\documentstyle{mn}

\newcounter{parentequation}

\def\ltsima{$\; \buildrel < \over \sim \;$}
\def\gtsima{$\; \buildrel > \over \sim \;$}
\def\simlt{\lower.5ex\hbox{\ltsima}}
\def\simgt{\lower.5ex\hbox{\gtsima}}

\def\etal{{et al.}~\rm}

\begin{document}

\title[Cosmic Microwave Background] {Principal Component Analysis of
the Cosmic Microwave Background Anisotropies: Revealing The Tensor
Degeneracy}

\author[G. Efstathiou]{G. Efstathiou\\
Institute of Astronomy, Madingley Road, Cambridge, CB3 OHA. \\
Theoretical Astrophysics, Caltech, Pasadena, CA 91125, USA.
}

\maketitle

\begin{abstract}
A principal component analysis  of cosmic microwave background
(CMB) anisotropy measurements is used to investigate degeneracies
among cosmological parameters.  The results show that a degeneracy
with tensor modes -- the `tensor degeneracy' -- dominates
uncertainties in estimates of the baryon and cold dark matter
densities, $\omega_b = \Omega_b h^2$, $\omega_c = \Omega_c
h^2$\footnotemark, from an analysis of CMB anisotropies alone. 
The principal component  analysis agrees
well with a maximum likelihood analysis of the observations,
identifying the main degeneracy directions and 
providing an impression of the effective dimensionality of the
parameter space.

\vskip 0.2truein

\noindent
{\bf Key words:} cosmic microwave background- cosmology:miscellaneous.

\vskip 0.3 truein
\end{abstract}\footnotetext{ Here $h$ is Hubble's constant $H_0$ in units of
$100{\rm km}{\rm s}^{-1} {\rm Mpc}^{-1}$}

\noindent


\section{Introduction}

Since the discovery of CMB anisotropies by the COBE team 
(Smoot \etal 1992) there has been rapid progress on the observational
front, culminating with publication earlier this year of evidence for
multiple acoustic peaks in the CMB power spectrum (Netterfield \etal
2001; Lee \etal 2001; Halverson \etal 2001). It has long been known
that accurate measurements of the CMB anisotropies can be used to
estimate parameters characterising the primordial fluctuations, the
geometry of the Universe and its matter content ({\it e.g.} Jungman
etal 1996; Bond, Efstathiou \& Tegmark 1997; Zaldarriaga, Spergel \&
Seljak 1997). In fact, the most recent measurements paint a
gratifyingly consistent picture compatible with the simplest models of
inflation ({\it i.e.}  a spatially flat Universe with scale-invariant
adiabatic fluctuations ).  Furthermore, the derived value of the
baryon density $\omega_b$ appears to be consistent with the value
$\omega_b = 0.020 \pm 0.002$ inferred from primordial nucleosynthesis
and deuterium abundance measurements from quasar absorption line
spectra (Burles, Nollett \& Turner 2001, and references therein).

It has also long been known that there are significant degeneracies
amongst cosmological parameters estimated from CMB anisotropies, {\it
i.e.} parameter combinations exist that produce nearly identical CMB power
spectra (Bond \etal 1994; Efstathiou \& Bond 1999, hereafter
EB99). The best known is the {\it geometrical degeneracy} between the
matter and vacuum energy densities, $\Omega_m$ and $\Omega_\Lambda$,
and the curvature $\Omega_k = 1 - \Omega_m - \Omega_\Lambda$. This 
degeneracy  is almost exact and precludes reliable estimates of either
$\Omega_\Lambda$ or the Hubble parameter $h$ from measurements of the
CMB anisotropies alone. The existence of parameter degeneracies means
that the best fitting parameters and their errors can be extremely
sensitive to the chosen parameter set ({\it e.g.} whether the Universe is
assumed to be spatially flat) and to adopted `prior distributions'
({\it e.g.} observational constraints on the Hubble constant). This
complicates the interpretation of CMB anisotropy parameter studies and
the intercomparison of limits determined by different authors.

Almost all CMB parameter analyses (with the noteable exception of the
work of Tegmark and collaborators, (Tegmark 1999; Tegmark, Zaldarriaga
\& Hamilton, 2001; Wang, Tegmark \& Zaldarriaga, 2001) have ignored
a tensor component (see for example, Lange \etal 2001; Jaffe \etal
2001; de Bernardis \etal 2001; Pryke \etal 2001; Stompor \etal 2001).
This special case is certainly interesting because a wide class of
inflationary models predict a negligible contribution from tensor
modes. Indeed it has been argued persuasively
that the absence of tensor modes is
generic to any model in which the inflaton potential is related to
the Higgs sector of a grand unified theory (Lyth 1997).

However, so little is known about inflation (if indeed inflation
occured) that it may be dangerous to neglect a tensor mode,
particularly if some cosmological parameters of interest are sensitive
to tensor modes. In fact, in single-field inflation models the
relative amplitudes of the tensor and scalar tensor modes ($r$) and
their spectral indices ($n_t$ and $n_s$) are related to the inflaton
potential and its first two derivatives ({\it e.g.} see Hoffman \&
Turner 2001 for a recent discussion). In this class of model the
relation between $n_s$ and $n_t$ is model dependent while the relation
between $n_t$ and $r$ depends only on the validity of the slow-roll
approximation. However, even the latter relation can be violated in multi-field
inflationary models. These relations may also be violated in some superstring
inspired models (see {\it e.g.} Lidsey, Wands \& Copeland 2000,
and references therein). Examples of the latter include the
pre-big bang model of Veneziano and collaborators ({\it e.g.}
Buonnono, Damour \&Veneziano, 1999) which it is argued can lead
to `blue' scalar spectral indices ($n_s >1$), and the ekpyrotic
scenario (Khoury, Ovrut, Steinhardt \& Turok 2001) which produces
a strongly blue tensor mode spectrum.

The point of view taken in this paper is to define a minimal parameter
set on the assumption that the primordial fluctuations are Gaussian,
adiabatic and featureless ({\it i.e.} defined by power-law spectral
indices). Thus the minimal model is specified by $9$ parameters: four
parameters specifying the amplitudes and spectral indices of scalar
and tensor components ($\bar Q$, $r_{10}$, $n_s$ and $n_t$, see Section 2
for more precise definitions), four parameters defining the matter
content and curvature of the Universe ($\omega_b$, $\omega_c$,
$\Omega_\Lambda$, $\Omega_k$) and a single parameter $\tau_{opt}$
quantifying the optical depth to Thomson scattering since the Universe
was reionized.  However, as pointed out by EB99, including a tensor
component with no assumed constraints between the parameters $n_s$,
$n_t$ and $r_{10}$ introduces an new major degeneracy between cosmological
parameters that we will call the {\it tensor degeneracy} in this
paper. This degeneracy has a dramatic effect on the permitted ranges
of some parameters, in particular the baryon and cold dark matter
densities $\omega_b$ and $\omega_c$.

One approach to these degeneracies is to apply brute-force maximum
likelihood analysis to a large parameter set (see the papers by
Tegmark and collaborators). Here we show that the observational data
has now improved to the point that a simple principal component
analysis of the Fisher matrix defined by the data identifies the major
degeneracy directions (geometrical and tensor) and provides a useful
first approximation to the likelihood function. Using the principal
components it is easy to analyse the correlations between physical
parameters introduced by the tensor degeneracy and to assess the
effects of introducing external (non-CMB) constraints on the
parameters.

\section{Principal component analysis}

We use the compilation of band power estimates $\Delta T_B^2$ and their
covariance matrix $C_{BB^\prime}$ computed by Wang \etal (2001,
hereafter WTZ01) from $105$ CMB anisotropy measurements. These band
power estimates include a model for calibration and beam errors (see
WTZ01 for further details). Each band power estimate is related to the
power spectrum $C_\ell$ of the CMB anisotropies by
\begin{equation}
 \Delta T_B^2 = {T_0^2 \over 2 \pi} \; \sum_\ell \ell(\ell+1) C_\ell
W_B(\ell) \label{2.1}
\end{equation}
where $W_B$ is the window function for each band power (also computed
by WTZ01).  These band-power estimates are plotted in Fig. 1. Also
plotted in this figure is a fiducial model\footnote{The CMB power
spectra in this paper were computed using the CMBFAST code of Seljak
\& Zaldarriaga (1996).}  with the following parameters:
$\omega_b=0.020$, $\omega_c = 0.13$, $h=0.7$ ($\Omega_b = 0.04$,
$\Omega_c = 0.26$), $\Omega_k=0$, $\tau_{opt}=0.1$, $n_s = 1$,
$n_t=0$, $r_{10}=0.2$. These parameters provide an extremely good fit to
the observations and are very close to the concordance values
determined by WTZ01 and the author from a full likelihood
analysis. The addition of a small tensor component has little effect
on the fit shown in Fig. 1, but is introduced to regularize the
Fisher matrix of equation (\ref{2.2}).

\begin{figure}

\vskip 3.0 truein

\includegraphics{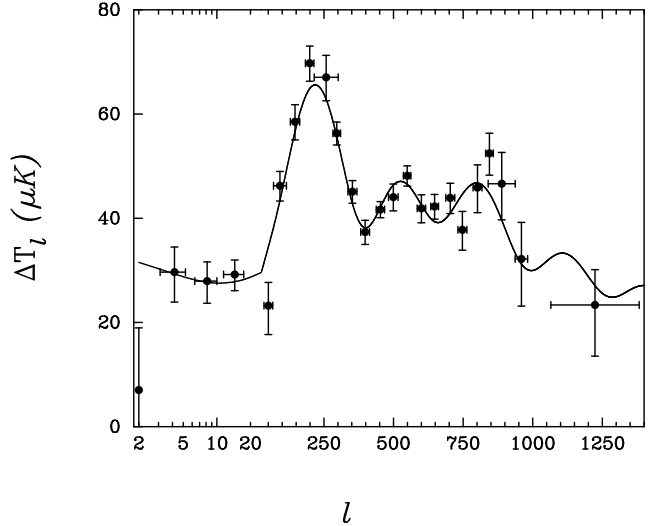}

\caption
{The points show band-averaged observational estimates of the CMB
power spectrum from WTZ01 plotted against multipole inde
using a log-linear abscissa. The error bars show
$\pm 1\sigma$ errors. The line shows the CMB power
spectrum for the fiducial inflationary model discussed in the
text.}
\label{figure1}
\end{figure}

\begin{figure*}

\vskip 3.7 truein

\includegraphics{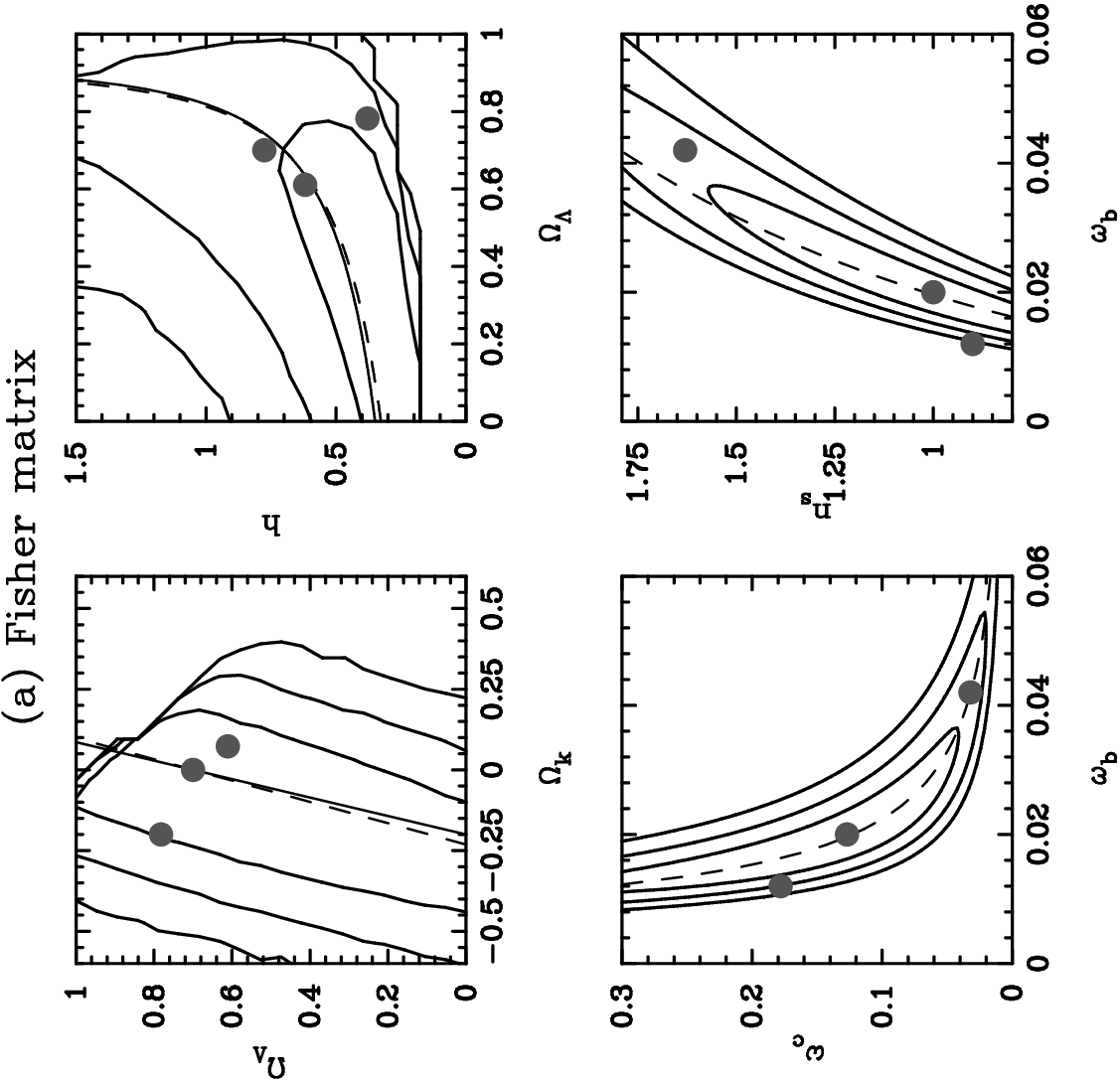}

\includegraphics{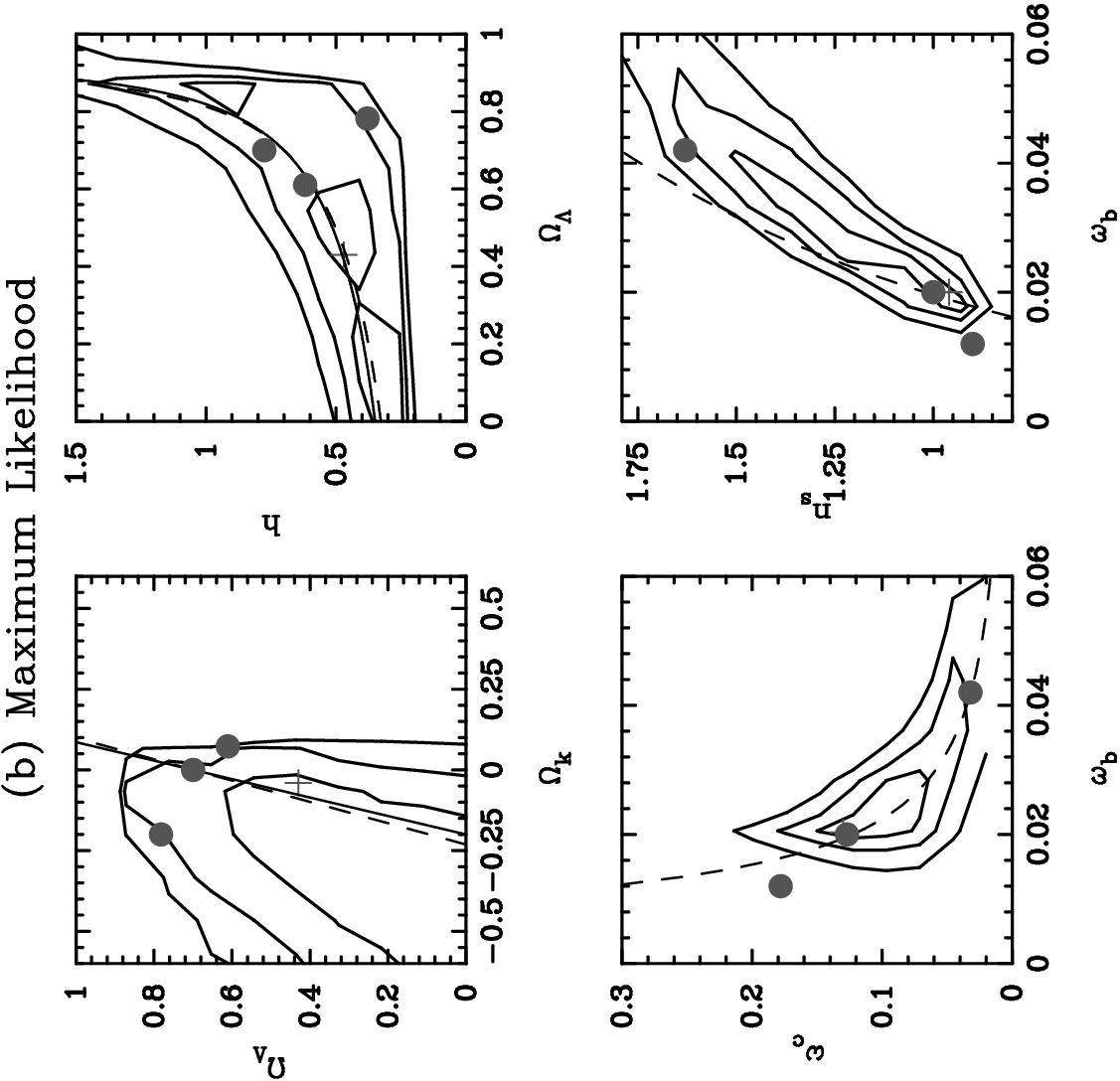}

\caption
{Various projections of the likelihood functions chosen to illustrate
the geometrical and tensor degeneracy. One, two and three-sigma
likelihood contours are plotted in each panel. The panels to the left
were computed from the Fisher matrix (equation \ref{2.2}) and the
panels to the right show `pseudo-marginalized' contours computed from
a full likelihood analysis of the data described by Efstathiou \etal
(2001). The solid lines in the upper panel show the approximate locus
of the geometrical degeneracy computed from the constraint $\Omega_D =
{\rm constant}$ and from the constraint equation (\ref{2.5}). The
dashed lines in the upper and lower panels show the degeneracy
directions defined by the component $X_9$ and $X_8$ respectively.  The
filled circles show the parameters of the nearly degenerate models
plotted in Fig. 3. The crosses in figures 2b show the positions of the
peaks in the likelihood function.}
\label{figure2}
\end{figure*}

Given the covariance matrix of the observations, we can form
the Fisher matrix for the (mean subtracted) parameter set $\{s_i\}$:
\begin{equation}
 F_{ij} =  
   \sum_{B B^\prime} C_{BB^\prime}^{-1} {\partial \Delta T^2_B
\over \partial s_i} { \partial \Delta T^2_{B^\prime} 
\over \partial s_j}.   \label{2.2}
\end{equation}
In defining the parameter set $s_i$, we use ${\rm ln} \omega_b$ and
${\rm ln} \omega_c$ rather than $\omega_b$ and $\omega_c$ and 
$\Omega_D = \Omega_k - 0.286\Omega_\Lambda$ rather than
$\Omega_k$. The latter expresses the geometrical degeneracy
(see EB99),
since if all parameters other than $\Omega_k$ and
$\Omega_\Lambda$  are held fixed,  the condition $\delta \Omega_D=0$
preserves the positions of the acoustic peaks for small variations of
the parameters about  those 
of the fiducial model. 
The normalization parameter $\bar Q$ is defined  following
EB99 so
that $\bar Q^2$ is the  mean band power ($\sum_{\ell < 1500}
\ell(\ell+1) C_\ell)$)
of a model relative to that of the fiducial model plotted
in Fig. 1. With this definition, $\bar Q$ is observationally
well constrained and is largely decoupled from variations in 
cosmology and  the optical depth $\tau_{opt}$ (unlike measures of
the amplitude related to low multipoles). The tensor to scalar
ratio $r_{10}$ is defined so that $C^T_{10} = r_{10}C^S_{10}$.
With these definitions, we can compute the Fisher matrix 
(\ref{2.2}) from derivatives of the power spectra $C^S_\ell$
and $C^T_\ell$, as in standard analyses of parameter forecasting
(Jungman \etal 1996, Bond \etal 1997). Alternatively, we could
compute the Hessian matrix by estimating second derivatives of
the likelihood function around the maximum likelihood value
either directly or by summing over first and second derivatives of
$\Delta T^2_B$. In most situationa
using either the Fisher and Hessian matrixes should give
closely similar results, and in fact, the second derivatives of 
$\Delta T^2_B$  are often ignored in order to stabilise numerical evaluations
of the Hessian (see {\it e.g.} Press \etal 1992).

Having computed the Fisher matrix, we diagonalise it,
\begin{equation}
{\bf  F} = {\bf U} {\bf \Lambda} {\bf U^T}, \qquad {\bf \Lambda} =
diag( \lambda_1, \lambda_2, \dots \lambda_N).   \label{2.3}
\end{equation}
The matrix ${\bf U}$ defines a set of principal
components, ${\bf X}$,  {\it i.e.} orthogonal linear combinations
of the original parameters,
\begin{equation}
{\bf X} = {\bf U}^T {\bf s},   \label{2.4}
\end{equation} 
such that the variance of the component $X_i$ is equal to $1/\lambda_i$.
(See BE99 for an application of principal component analysis to
the MAP and Planck satellites\footnote{Descriptions of these  satellites
can be found on the following web pages: 
{http://astro.estec.esa.nl/SA-general/Projects/Planck} and
{http://map.gsfc.nasa.gov} }.

\begin{table*}
\bigskip
\centerline{\bf Table 1: Principal components for data of Fig. 1}
\begin{center}
\begin{tiny}
\begin{tabular}{ccccccccccc} \hline \hline

 & $\lambda_i^{-1/2}$ & ${\rm ln}\omega_b$ & {\rm ln} $\omega_c$ & $n_s$ &  $\bar Q$      &  $\tau_{opt}$     
& $\Omega_\Lambda$      & $\Omega_D$ & $n_t$ & $r_{10}$ \cr

$X_1$ &   $\;\;$8.4923E-03 &  $\;\;$8.0062E-02 & -1.8131E-01 &  $\;\;$2.0836E-02 & -1.5648E-01 & -6.8989E-03& -3.0498E-03 &  $\;\;$9.6731E-01 &  $\;\;$3.1160E-03 & -6.4147E-03 \cr
$X_2$ &   1.6231E-02 &  $\;\;$2.9071E-02 & -3.1679E-02 & -3.7260E-01 &  $\;\;$9.0306E-01 &  $\;\;$1.0592E-01&  $\;\;$6.3355E-02 &  $\;\;$1.4733E-01 & -4.5002E-02 &  $\;\;$6.9623E-02 \cr
$X_3$ &   3.1626E-02 & -1.5949E-01 & -1.0387E-01 &  $\;\;$8.0645E-01 &  $\;\;$3.9347E-01 & -2.9357E-01& -1.8037E-01 &  $\;\;$3.5921E-02 &  $\;\;$1.1322E-01 & -1.6053E-01 \cr
$X_4$ &   7.1340E-02 & -7.3973E-01 &  $\;\;$1.7199E-01 &  $\;\;$1.7869E-01 & -1.1711E-02 &  $\;\;$4.2868E-01&  $\;\;$3.4928E-01 &  $\;\;$9.3853E-02 & -1.7909E-01 &  $\;\;$2.1096E-01 \cr 
$X_5$ &   1.3033E-01 &  $\;\;$5.4011E-01 &  $\;\;$6.8276E-01 &  $\;\;$3.2525E-01 &  $\;\;$7.0219E-02 & $\;\;$2.4081E-01&  $\;\;$2.1371E-01 &  $\;\;$9.0979E-02 & -1.0073E-01 &  $\;\;$9.6515E-02 \cr
$X_6$ &   2.8274E-01 & -2.1034E-01 &  $\;\;$4.2130E-01 & -1.2216E-01 & -9.8231E-03 & -4.0134E-01& -5.1452E-01 &  $\;\;$9.6988E-02 & -9.6036E-02 &  $\;\;$5.6463E-01 \cr
$X_7$ &   8.9800E-01 &  $\;\;$2.1023E-01 & -3.8577E-01 &  $\;\;$1.4257E-01 & -1.7712E-03 & -3.3061E-01&  $\;\;$4.5135E-01 & -8.9152E-02 & -4.5136E-01 &  $\;\;$5.1182E-01 \cr
$X_8$ &   1.5724E+00 &  $\;\;$1.9488E-01 & -3.5449E-01 &  $\;\;$1.8691E-01 &  $\;\;$3.5187E-03 &  $\;\;$5.7374E-01& -3.6421E-01 & -8.0544E-02 &  $\;\;$2.6539E-01 &  $\;\;$5.1250E-01 \cr
$X_9$ &   3.8197E+00 & -4.4314E-02 &  $\;\;$8.7287E-02 & -4.9695E-02 & -1.3832E-03 & -2.4747E-01&  $\;\;$4.4122E-01 &  $\;\;$1.9666E-02 &  $\;\;$8.1212E-01 &  $\;\;$2.6851E-01 \cr
$\langle s_i^2 \rangle^{1/2}$  & & $0.41$ & $0.75$ & $0.38$ & $0.023$ & $1.3$ &
$1.8$ & $0.17$ & $3.1$ & $1.4$ \cr
 \hline
\end{tabular}
\end{tiny}
\end{center}
\end{table*}

The eigenvalues and components of ${\bf U}^T$ for our chosen set of
variables and fiducial model are listed in Table 1. The principal
components have been ordered by their expected variance so that $X_1$
is the best determined parameter and $X_9$ is the worst. The last line
in the table lists the predicted variances of the parameters $s_i$,
($\sqrt{(F^{-1}_{ii})}$).  As can be seen from Table 1, the present data
can be used to constrain three parameter combinations well and 
three extremely
poorly, with the remaining constrained at intermediate levels of
accuracy. The best and worst determined principal components have a
straightforward interpretation. $X_1$ has a high weight from
$\Omega_D$, and so, in effect, measures the positions of the acoustic
peaks; $X_2$ has a high weight from the overall amplitude of the
spectrum $\bar Q$; $X_3$ has high weights from the scalar spectral
index $n_s$ and $\tau_{opt}$ and provides a measure of the shape of
the fluctuation spectrum.  The component $X_9$ accounts for almost all
of the variance of $\Omega_\Lambda$, $n_t$ and $r_{10}$,  and describes the
geometrical degeneracy and the extremely poor constraints that the
present data place on tensor modes. The component $X_8$ 
describes the tensor degeneracy and accounts for
the almost all of the variance of the parameters ${\rm ln} \omega_b$,
${\rm ln} \omega_c$, $n_s$ and $ \Omega_D$. Together, the
components $X_9$ and $X_8$ accounts for almost all of the variance of
all of the parameters with the exception of the amplitude $\bar Q$.

The principal component analysis therefore suggests that 
acceptable models lie on a plane within the $9$ dimensional parameter
space. This is illustrated in Fig. 2a which shows likelihood contours
in  projections  chosen to  illustrate the geometrical and tensor degeneracies.
The upper panels show the geometrical degeneracy in the parameter pairs
$\Omega_\Lambda-\Omega_k$ and $h - \Omega_\Lambda$. The dashed lines
show the degeneracy directions defined by the component $X_9$. The
solid line in the $\Omega_\Lambda - \Omega_k$ plane shows the constraint
$\delta \Omega_D = 0$. The solid
line in the $h - \Omega_\Lambda$ plane is computed from the constraint
equation
\begin{equation}
 h = {(\omega_b + \omega_c)^{1/2} \over ( 1 - \Omega_k - \Omega_\Lambda)^{1/2}}
 = { (\omega_b + \omega_c)^{1/2} \over \left ( 1 - \Omega_D - {4.5 \over 3.5} 
\Omega_\Lambda\right )^{1/2}},  \label{2.5}
\end{equation}
with $\omega_b$, $\omega_c$ and $\Omega_D$ set to the values of the target
model. The dotted lines in the lower two panels show the degeneracy directions
defined by the component $X_8$.

The Fisher matrix and the associated principal component analysis is
approximate and it is not obvious {\it a priori} how well they decribe
degeneracies especially for parameter values that are quite a long way
from those of the target model.  Fig. 2b shows the analogous
contours computed from a full likelihood analysis of the data of Fig.
1 (see the companion paper of Efstathiou \etal 2001
for a detailed discussion).
The agreement is surprisingly good. The general directions of the
geometrical and tensor degeneracies follow those of the Fisher matrix
analysis. There are some differences however. The low order CMB multipoles
offer some discrimination of models with high values of
$\Omega_\Lambda$ via the integrated Sachs-Wolfe effect (see EB99),
thus models with $\Omega_\Lambda \simgt 0.88$ are excluded by the
data. The maximum likelihood analysis shows that the tensor degeneracy
allows high values of $\omega_b$ but that the likelihood function
falls sharply for $\omega_b \simlt 0.018$ (interestingly only just
below the value favoured from primordial nucleosynthesis
deuterium constraints).

\begin{figure*}

\vskip 3.4 truein

\includegraphics{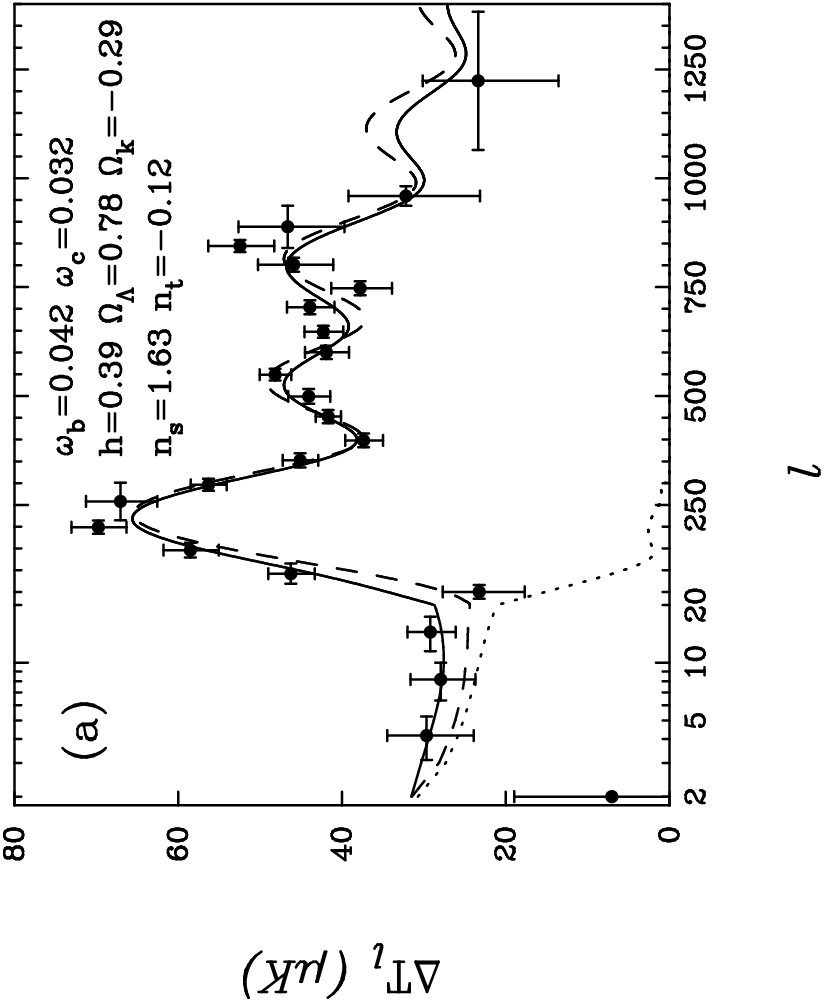}

\includegraphics{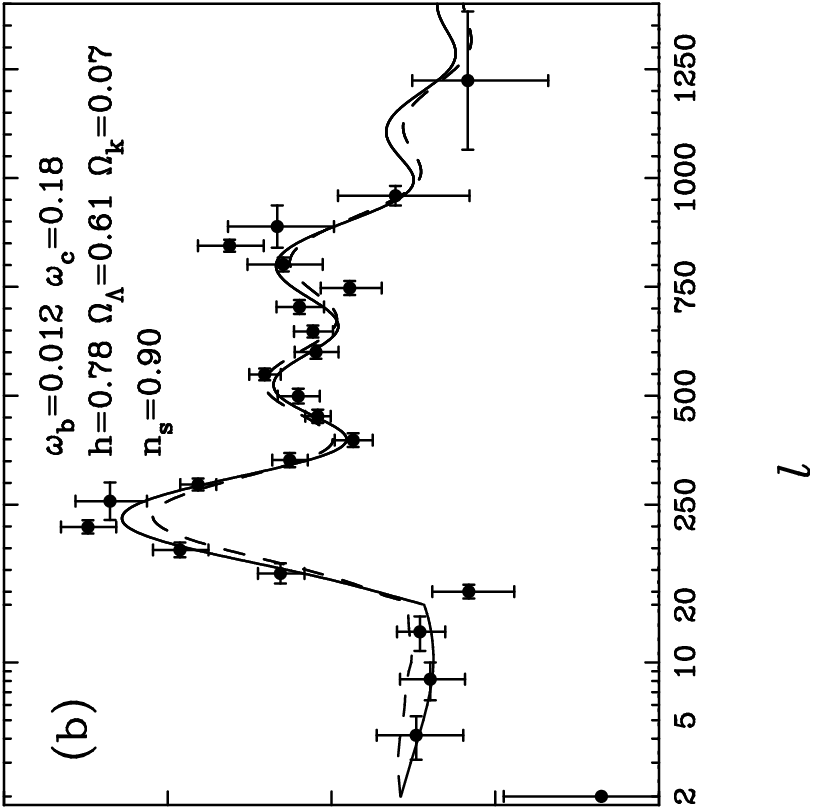}

\caption
{The data and fiducial model (solid line) of Fig. 1. The dashed
lines show CMB power spectra for nearly degenerate models with a
high baryon density (Fig. 3a) and a low baryon density (Fig. 3b)
chosen to lie along the direction of the tensor degeneracy. The dotted
line in Fig. 3a shows the contribution of the tensor  component.
The low model with low baryon density in Fig. 3b has a neglible
contribution from tensor modes.}
\label{figure3}
\end{figure*}

 The three filled circles in Fig. 2 show the parameter values for the
target model and for two nearly degenerate models with extreme values
of $\omega_b$ chosen to lie along the tensor degeneracy
direction. (The device used here, of varying low order principal
components to produce degenerate pairs of models, was used by EB99 to
investigate parameter degeneracies for the MAP and Planck satellites.)
One model has $\omega_b = 0.042$ and the other has $\omega_b =
0.012$. The models were intentionally chosen to have similar values
of $\Omega_\Lambda$,  consequently the high baryon density model
has a low value of $h$ which one might argue is incompatible with
independent measurements of the Hubble constant (Freedman \etal 2001).
However, our purpose here is to show how the constaints on $\omega_b$
derived from CMB anisotropies {\it alone} are weakened when tensor modes
are included.
The high and low baryon density models are compared to the data and to the fiducial model in
Fig. 3.

Despite the very different parameter values, the models produce almost
identical CMB power spectra by construction. The exact likelihood
analysis (Fig. 2b) shows that the model with the high baryon density
is compatible with the data at about the $2 \sigma$ level. The low
baryon density model is formally excluded by the data at a high level
of significance ($> 3 \sigma$) because it fails to match the height of
the first acoustic peak. This is a characteristic feature of models
with a low baryon density. Nevertheless, the diagram is interesting because it
shows that that the lower limits on $\omega_b$ are extremely sensitive
to any residual systematic errors that might affect the peak heights
(see {\it e.g.} de Bernardis \etal 2001, figure 1).

\section{Discussion}

The reader might question the usefulness of the results presented in
the previous section. Firstly, the principal component analysis
provides only an approximate description of the parameter
degeneracies, whereas they emerge precisely from a brute force maximum
likelihood analysis.  Secondly, models at the extreme ends of the
ranges allowed by the tensor degeneracy have unusual
parameters (for example, the model with $\omega_b=0.042$ in Fig. 3b
has a low Hubble constant and a high value of the scalar spectral
index) and so are surely excluded by other observational
constraints. We discuss each of these points in turn.

\vskip 0.1 truein

\noindent
{\it (i) Effective dimensionality:} The main use of the principal
component analysis is to assess the effective dimensionality of the
space of acceptable models within the multidimensional space defined
by the physical parameters $s_i$ and to assess whether this effective
dimensionality is sensitive to changes of the parameter set. For
example, let us assume that we are interested in the values of the
parameters $\omega_b$ and $\omega_c$. The principal component analysis
tells us that most of the variance of these parameters is contributed
by only two poorly constrained principal components. The values of
these parameters are therefore affected by major parameter
degeneracies which can only be removed by imposing external
constraints or by performing a fundamentally different type of CMB
exeperiment.  For example, the effects of the tensor degeneracy on
$\omega_b$ and $\omega_c$ can be broken by extending the CMB
measurents to much higher multipoles (see EB99) and/or by setting
limits on a tensor component from an analysis of a B-type polarization
pattern in the CMB (see {\it e.g.} Kamionkowski \& Jaffe, 2000, and
references therein).

\vskip 0.1 truein

\noindent
{\it (ii) Complementary information:} CMB parameter degeneracies can
be broken by invoking complementary information. A well known example
is the combination of Type Ia supernovae measurements with the CMB to
break the geometrical degeneracy (see {\it e.g.} de Bernardis \etal
2001 for a recent analysis). WTZ01 break the tensor degeneracy by
combining the CMB data with a number of constraints including
estimates of the power spectrum on small scales from observations of
the Ly$\alpha$ forest (Croft \etal 2001) and limits on the Hubble
constant from HST Hubble Key Project (Freedman \etal 2001). The
problem here is that it becomes progressively more difficult to assess
whether parameter values are affected by systematic errors as more
external constraints are applied, particularly if the external
constraints involve complex observations and assumptions.
(For example, how can we test empirically whether density
fluctuations in the  inter-galactic medium 
as traced by the Ly$\alpha$ forest match those of the dark matter?).
 Even if the
parameter values are shown to remain consistent as external constraints
are applied, there is no guarantee that the final combined likelihood
distribution will be accurate. Ideally, we would like to apply as
small a number of external constraints as possible using data sets
with well controlled errors. The Fisher matrix analysis of the
previous section offers a guide as to which external constraints will
be most effective at breaking 
degeneracies.
The effects of external constraints can easily be assessed
using the Fisher matrix: let $P_{ij}$ be the covariance matrix
of the parameters $s_i$ from external constraints,  then the 
covariance matrix
after combining with the CMB is  
$$
C_{ij} = (F_{ij} +
P_{ij}^{-1})^{-1}.
$$ 
As menstioned in Section 2, the models plotted in Fig. 3 were
chosen by construction to have similar values of $\Omega_\Lambda$,
so it is clear that combining the CMB results with constraints on
$\Omega_\Lambda$ from Type Ia supernovae observations will have little
effect on the parameters $\omega_b$ and $\omega_c$. Constraints on
the Hubble constant will be more effective in 
narrowing the range of allowed values of $\omega_b$ and
$\omega_c$. However, from Fig. 2 we can see that the tensor
degeneracy exhibits a strong correlation between $\omega_b$,
$\omega_c$, and the scalar spectral index $n_s$. Even a relatively poor
constraint on the scalar spectral index will lead to a sharp
tightening of the $\omega_b$ and $\omega_c$ error contours. (A
constraint $\Delta n_s$ on the spectral index will narrow the
error on the baryon density to  
$\Delta {\rm ln} \omega_b \approx (U_{18}/U_{38}) \Delta n_s$, 
{\it i.e.} $\Delta {\rm ln} \omega_b \approx 1.04 \Delta n_s$ .) This
suggests that the recent estimates of the galaxy power spectrum from
the 2dF Galaxy Redshift Survey (Percival \etal 2001) will be highly
effective at breaking the tensor degeneracy,  leading to tight
constraints on the matter content of the Universe. This is
borne out by a detailed analysis presented in a companion paper
(Efstathiou \etal 2001).

\vskip 0.2 truein

\noindent
{\bf Aknowledgments:} I thank Caltech for the award of
a Moore Scholarship. I thank Andrew Liddle for discussions of
inflationary models and various members of the 2dF Galaxy Survey
team for their comments on an earlier draft.

\end{document}